\documentclass[twocolumn,
               superscriptaddress,
               floatfix,
               longbibliography, 
               apl]{revtex4-2}  
               
\usepackage{graphicx} 
\usepackage{bm}	
\usepackage{color}                     
\usepackage{xcolor}
\usepackage{epsfig}
\usepackage{amsmath} 
\usepackage{amssymb} 
\usepackage{longtable} 
\usepackage{float}
\usepackage{mathtools}
\usepackage{xparse}
\usepackage{hyperref}
\usepackage{times}
\usepackage[normalem]{ulem}
\usepackage{sidecap}
\usepackage{appendix}
\usepackage{comment}
\usepackage{placeins}

\newcommand{\eg}{\emph{e.g.}, }
\newcommand{\ie}{\emph{i.e.}, }
\newcommand{\gqoc}{family-control}
\newcommand{\paramstgt}{\alpha}
\newcommand{\gcontrolfunc}{f_{\paramstgt}(t)}
\newcommand{\bra}[1]{\langle #1 |}
\newcommand{\ket}[1]{|#1\rangle}

\begin{document}

\title{
Optimal control of families of quantum gates
}
\author{Fr\'ed\'eric Sauvage}
\author{Florian Mintert}

\affiliation{Physics Department, Blackett Laboratory, Imperial College London, Prince Consort Road, SW7 2BW, United Kingdom}
\date{\today}

\begin{abstract}
Quantum Optimal Control (QOC) enables the realization of accurate operations, such as quantum gates, and support the development of quantum technologies.
To date, many QOC frameworks have been developed but those remain only naturally suited to optimize a single targeted operation at a time.
We extend this concept to optimal control with a continuous family of targets,
and demonstrate that an optimization based on neural networks can find families of time-dependent Hamiltonians that realize desired classes of quantum gates in minimal time.
\end{abstract}

\maketitle

After concerted efforts in the development of synthetic quantum systems we have access to a variety of systems with sufficiently long coherence time to perform a series of coherent operations.
In the community's effort to turn such systems into technological applications,
Quantum Optimal Control (QOC)~\cite{Werschnik_2007,Glaser2015} helps to increase the precision and rate of desired operations.
Common problems successfully addressed by means of QOC include
the realization of quantum gates or entangled states in few-body or many-body systems~\cite{PhysRevA.77.052334,Dolde2014,PhysRevLett.112.190502,Heeres2017,Schafer2018,PhysRevLett.122.110501,PhysRevLett.123.170503,Omran570,PhysRevX.10.021058} and the refinement of metrology protocols~\cite{PhysRevX.8.021059,doi:10.1116/5.0006785}.

Current tasks of optimal control are mostly focused on the realization of a single target operation, such as the preparation of one specific state or the implementation of one specific gate.
Yet, as quantum technologies mature, it becomes important to enlarge the range of operations which can be accurately implemented on a device.
For instance, in the context of Noisy-Intermediate Scale Quantum (NISQ) devices~\cite{preskill2018quantum}, 
augmenting the set of available elementary gates allows for more compact compilation of quantum circuits, \ie their decomposition into these elementary gates.
Already the inclusion of continuous families of two-qubit gates to a typical gate-set, composed of single-qubit rotations and a two-qubit entangling gate, can lead to a significant reduction in gate count~\cite{PhysRevLett.125.120504, PRXQuantum.1.020304, Abrams2020}.
This, in turn, opens the possibility to run more expressive computations before the onset of decoherence, a key limitation in current technology.
That is, the ability to implement a broader range of optimized operations has the potential to substantially increase the utility of current quantum hardware.

\begin{figure}
	\includegraphics[width=0.49\textwidth]{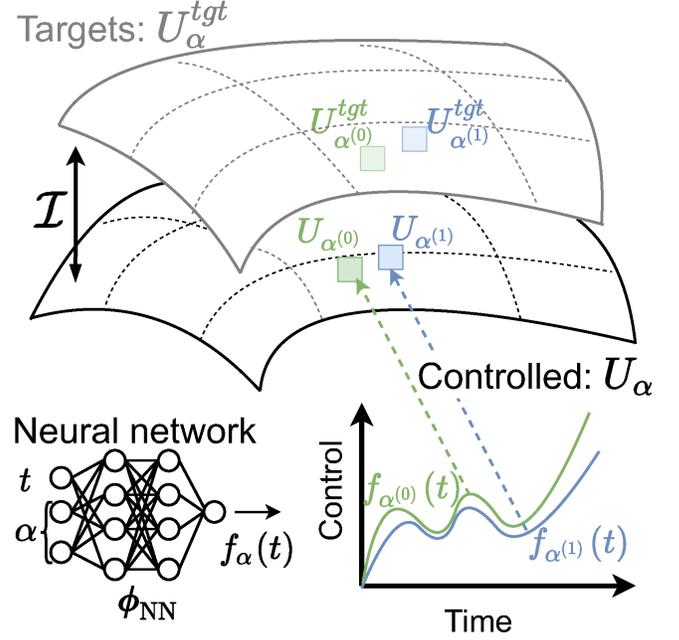}
	\caption{
    Optimal control of a continuous family of target gates $U^{tgt}_{\alpha}$ indexed by the target parameter $\alpha$ which can be either a scalar or a vector.
    The time-dependent controls $\gcontrolfunc$ which now also depend on $\alpha$ are modeled by a neural network~(NN).
    This NN effectively parameterizes a continuous family of controlled gates $U_{\alpha}$ where each point corresponds to the propagator obtained by evolving the system in time (dashed arrows) under the controls produced by the NN (illustrated for two set of target parameters $\alpha^{(0)}$ and $\alpha^{(1)}$).
    Training of the framework consists in optimizing the weights of the NN, denoted $\phi_{NN}$, such that the deviation $\mathcal{I}$ between the controlled and target families of operations is minimized.
	}
	\label{fig:gqoc_pres}
\end{figure}

Despite the many flavours of QOC frameworks that have been proposed (e.g.,~\cite{KHANEJA2005296,PhysRevA.84.022326,doi:10.1063/1.3691827,PhysRevLett.114.200502,PhysRevLett.118.150503,PhysRevA.95.042318,PhysRevLett.120.150401,PhysRevX.8.031086,Niu2019,PRXQuantum.1.020322,PRXQuantum.2.020332}), it remains the case that current methodologies are only naturally suited to consider a single control task at a time.
We thus aim at lifting the original scope of QOC from the control of a single target operation to  the control of continuous families of targets.
This is achieved with a neural-network~(NN) modelling the dependency between Hamiltonians to be engineered and control tasks to be solved.
Efficient training of the framework by means of gradient-descent is facilitated by recent advances in the field of automatic-differentiation~\cite{NEURIPS2018_69386f6b}. 
Such framework, dubbed \gqoc, is sketched in Fig.~\ref{fig:gqoc_pres} and is now explained in detail.

Typically, the central task in QOC is the identification of the time-dependent Hamiltonian $H(t)$ that induces a propagator $U(t)$ with desired properties.
This is formulated in terms of a cost functional $\mathcal{I}(H(t))$ that is to be minimized.
A common example would be the task of realizing a target controlled--not gate $U^{tgt}=\ket{0}\bra{0}\otimes I+\ket{1}\bra{1}\otimes \sigma_x$ at a given time $T$, with the cost $\mathcal{I}(H(t))=||U(T)-U^{tgt}||$ being a measure of the deviation between the controlled and target propagators.
A corresponding task of \gqoc\ could be the realization of the family of target gates $U^{tgt}_{\alpha}=\ket{0}\bra{0}\otimes I+\ket{1}\bra{1}\otimes\exp(-i\alpha \sigma_x)$ with a variable angle $\alpha$.
In this case, the overall task to be solved would be the identification of a continuum of Hamiltonians $H_\alpha(t)$, parameterized by the angle $\alpha$, such that any of the propagators $U_\alpha=U_\alpha(t=T)$ induced by $H_\alpha(t)$ approximates the gate $U^{tgt}_{\alpha}$ as well as possible at $t=T$.
The corresponding functional would thus become $\mathcal{I}(H_\alpha(t))=\langle ||U_{\alpha}-U^{tgt}_{\alpha}||\rangle_{\alpha}$ with an average over $\alpha$.

The general formulation of such a \gqoc\ problem can be given in terms of the individual costs  $\mathcal{I}_\alpha(H_\alpha(t))$, where the target parameter $\alpha$ can be a single scalar or a vector.
The overall task to be solved is the identification of the continuum of time-dependent Hamiltonians $H_\alpha(t)$ that minimizes the averaged cost $\mathcal{I}=\langle \mathcal{I}_\alpha(H_\alpha(t))\rangle_\alpha$.

In principle, this can be addressed in terms of several control problems to be solved separately 
for a discretized set of target values $\{ \alpha^{(i)} \}$,and, an additional step of interpolation for any new target with $\alpha \notin \{ \alpha^{(i)} \}$.
Hardly any control problem, however, has a unique solution, or at least a unique solution that can be found in practice.
That is, there is no guarantee for two Hamiltonians $H_{\alpha^{(1)}}$ and $H_{\alpha^{(1)}}$ identified as optimal for similar values of $\alpha^{(1)}$ and $\alpha^{(2)}$, to be themselves similar.
Any attempt to find an optimal Hamiltonian for a value of $\alpha$ between $\alpha^{(1)}$ and $\alpha^{(2)}$ in terms of an interpolation between $H_{\alpha^{(1)}}$ and $H_{\alpha^{(2)}}$ can thus result in a Hamiltonian that utterly fails to realize the desired task.
In order to avoid this issue, it is desirable to require $H_{\alpha}$ to depend smoothly on $\alpha$.
Such requirement can be realized by means of an appropriate parameterization of the dependence of $H_{\alpha}(t)$ on both $\alpha$ and $t$.
Given that neural networks (NNs) provide the flexible structure to approximate multivariate continuous functions up to any desired precision~\cite{HORNIK1991251}, these are deemed ideally suited for the task at hand.

Time-dependence in a Hamiltonian is typically realized in terms of temporally modulated electro-magnetic fields that appear as one or several control functions $\gcontrolfunc$ in the Hamiltonian.
The scope of the NN is thus to model these functions.
To this intent, the parameters $\alpha$ and the time $t$ are taken to be the inputs of the NN, and the control values $\gcontrolfunc$ its outputs.
An example of such a NN is illustrated in Fig.~\ref{fig:gqoc_pres} for the case of two--dimensional parameters $\alpha$ and a single control function $f_{\alpha}(t)$; it is readily adapted to arbitrary dimensions of the parameters and number of controls by varying the sizes of the inputs and outputs accordingly.

The optimization -- \ie training of the neural network -- can be achieved with a variety of techniques, but gradient-descent training has the advantage of simplicity and scalability to high-dimensional problems.
Since the propagators $U_\alpha$ induced by the Hamiltonians $H_\alpha$ typically need to be constructed numerically, efficient means to take derivatives with respect to the control functions $\gcontrolfunc$ are essential.
Recent advances in the field of automatic differentiation~\cite{NEURIPS2018_69386f6b} give access to efficient differentiation over numerical solvers of differential equations.
This allows to combine seamlessly gradients over the evolution of the system
and over the weights of the NN.
Finally, even though the evaluation of the averaged cost $\mathcal{I}$ would always be based on a sum over discrete values of $\alpha$ rather than a proper integral, the output of the neural network is still continuous in $\alpha$,
and choosing different random sampling points at each step in the training process avoids finding solutions with artefacts resulting from the sampling.
Implementation details can be found in Sec.~I of Supp. Matt.

The following discussion exemplifies the framework sketched so far,
with the realization of quantum gates induced by the $n$--qubit Hamiltonian
\begin{equation}
\label{eq:H_nq}
\mathcal{H}_\alpha(t)=  \sum_{i < j=1}^{n} f_{\alpha}^{ij}\sigma_x^{(i)}\sigma_x^{(j)} + \sum_{i=1}^{n}  f_{\alpha}^{iy}\sigma_y^{(i)} + f_{\alpha}^{iz} \sigma_z^{(i)}\ ,
\end{equation}
with pairwise $\sigma_x \sigma_x$ interactions, single-qubit Pauli $\sigma_y$ and $\sigma_z$ terms,
and time-dependent control functions $f_{\alpha}^{ij}(t)$, $f_{\alpha}^{iy}(t)$ and  $f_{\alpha}^{iz}(t)$.
The Hamiltonian is sufficiently general so that any desired $n$--qubit unitary can be realized~\cite{schirmer2001complete},
but bounded control amplitudes $\{f_{\alpha}^{ ij},f_{\alpha}^{iy},f_{\alpha}^{iz}\}\in[-1,1]$
typically result in a finite minimal time required to realize a given unitary. 
Deviations between controlled and target gates are characterized in terms of the gate infidelities
\begin{equation}
\label{eq:loss_onequbit}
{\cal I}_{\paramstgt}(H_{\alpha}(t)) = 1 - \frac{1}{2^n}\left| \mbox{Tr}\ [U_{\paramstgt}^{\dagger} U^{tgt}_{\paramstgt}] \right|^2
\end{equation}
in the subsequent examples.

\begin{figure} 
	\includegraphics[width=0.49\textwidth]{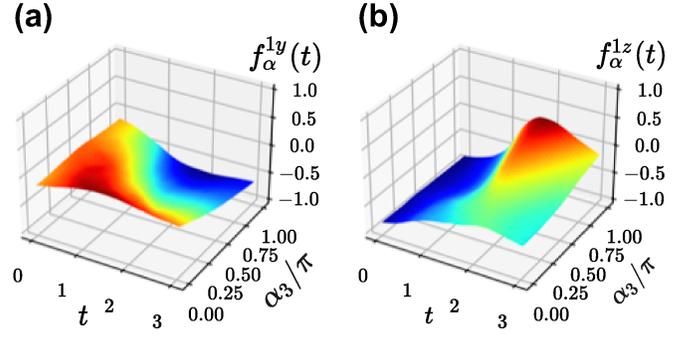}
	\caption{
	Control of the family of arbitrary single--qubit rotations $U^{1}_{\paramstgt}$ defined in Eq.~\eqref{eq:r1q} with the Hamiltonian given in Eq.~\eqref{eq:H_nq}.
	The framework is successfully trained to implement any of the target rotations, resulting in an average infidelity of $\bar{\mathcal{I}}=2 \times 10^{-4}$ (which is assessed based on new targets that were not seen during training).
	To visualize the controls produced by the NN, $\alpha_1$ and $\alpha_2$ are kept fixed to a value of $3 \pi /4$ and $\alpha_3 $ is varied in the range $[0,\pi]$.  
	The amplitudes of the two control fields $f_{\paramstgt}^{1y}$ (inset a), and $f_{\paramstgt}^{1z}$ (inset b) produced by the NN are plotted as a function of both the time $t$ and $\alpha_3 / \pi$.
	}
	\label{fig:1q_res}
\end{figure}

The basic workings of the present framework can be illustrated with the task of realizing the manifold of single-qubit gates
\begin{equation}
\label{eq:r1q}
U^{1}_{\paramstgt} = \exp\left(-i\frac{\alpha_1}{2}\sigma_z\right)\exp\left(-i\frac{\alpha_2}{2}\sigma_y\right)\exp\left(-i\frac{\alpha_3}{2}\sigma_z\right)
\end{equation}
for the three-dimensional target parameters $\alpha$ with components $\alpha_{j=1,2,3} \in [0, \pi]$, given the single-qubit version of $\mathcal{H}_{\alpha}(t)$ in Eq.~\eqref{eq:H_nq} and a fixed gate time $T=\pi$.
Here, and in all subsequent examples, the training stage is limited to $400$ iterations, 
with each iteration corresponding to an 
average of the gate infidelity in Eq.~\eqref{eq:loss_onequbit} taken over $128$ values of the parameters $\alpha$ uniformly sampled.

After training, the average gate infidelity $\bar{\cal I}=\langle{\cal I}_\alpha\rangle_\alpha$ resulting from the controls identified as optimal by the framework, is evaluated on an ensemble of $250$ random values of the parameters $\alpha$.
Crucially, this average is taken with respect to new parameter values (that is, corresponding to targets which have not been seen during training), and thus probes the ability of the framework to realize any gates belonging to the targeted family.
In this example, the average infidelity is as low as $\bar{\mathcal{I}}=2[3]\times 10^{-4}$, where the number in brackets indicates the standard deviation of the distribution.
Fig.~\ref{fig:1q_res} depicts the two control functions $f_{\alpha}^{1y}(t)$ and $f_{\alpha}^{1z}(t)$ (insets (a) and (b) respectively) as function of $\alpha_3/\pi$ and $t$ for $\alpha_1=\alpha_2=3\pi/4$, substantiating that the solutions produced by the neural network are indeed well-behaved, continuous functions of both the parameters $\alpha$ and time $t$.

\begin{figure}[t]
	\includegraphics[width=0.49\textwidth]{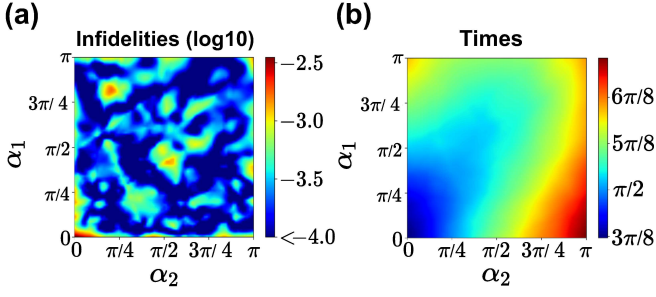}
	\caption{
	Infidelities and times for the family $U_{\paramstgt}^1$ of rotations. 
	In this example, both the (target-dependent) control functions and times are learnt concurrently.
	For the sake of visualization, results are plotted for a two-dimensional subset of the three--dimensional family of targets, with fixed parameter $\alpha_3=3\pi/4$, and discretized $\alpha_1, \alpha_2 \in [0, \pi]$ over a grid of $75 \times 75$ regularly spaced points.
	For each of the corresponding target gates a heat-map indicates the values of the infidelities $\mathcal{I}_{\alpha}$ achieved (inset a, in logarithmic scale) and the control times $T_{\alpha}$ entailed by the framework (inset b).
	}
	\label{fig:1q_res_trainableT}
\end{figure}

In addition to an optimization of control functions, the present framework is also well suited 
to identify a minimal gate time $T$, or even minimal target-dependent gate times $T_\alpha$ (Sec.~I~B  of  Supp. Matt.).
The latter is achieved by introducing a second neural network with the gate time $T_\alpha$ as an output.
Given the new cost ${\cal I}_\alpha + \mu \times T_\alpha$ comprised of the gate infidelity and the gate time as a penalty weighted with a scalar factor $\mu>0$,
this second neural network can be trained similarly to the case discussed above. 

In Fig.~\ref{fig:1q_res_trainableT} are reported the results from such an optimization with a small value $\mu=10^{-2}$ of the weight factor which is suitable to find high-fidelity gates close to the minimally required time.
Inset~(a) depicts the infidelity ${\cal I}_\alpha$ of the resulting gates for $\alpha_3=3\pi/4$ as function of $\alpha_1$ and $\alpha_2$.
Typical values are smaller than $10^{-3}$ and the average infidelity with the average taken over all three components of $\alpha$ is $\bar{\cal I}=4[5]\times 10^{-4}$.

Inset~(b) depicts the target-dependent minimized gate times, with again a fixed value of $\alpha_3=3\pi/4$ but varied $\alpha_1$ and $\alpha_2$.
The shortest gate time is obtained for $\alpha_1=\alpha_2=0$ (\ie for the target parameters $\alpha^{(0)}=[0,0,3\pi /4]$) in which case the constant control amplitudes $f_{\alpha^{(0)}}^{1z}(t)=1$ and $f_{\alpha^{(0)}}^{1y}(t)=0$
induce the desired gate $U^1_{\alpha^{(0)}}=\exp(-i\frac{3\pi}{8}\sigma_z)$ after a time $T_{\alpha^{(0)}}=3\pi/8$.
The obtained gate times grow with increasing values of $\alpha_1$ and $\alpha_2$, but always remain below the value of $\pi$ used in the above example.

While the ability to realize single-qubit gates is of substantial practical value, it is certainly not the challenging control problem that helps to demonstrate the actual strength of the framework.
This is better achieved in terms of two-qubit and three-qubit gates that are building blocks of quantum algorithms or digital quantum simulations.

\setlength{\tabcolsep}{0pt}
\begin{table}[htb]
\caption{Control under the Hamiltonian of Eq.~\eqref{eq:H_nq} of families of $n=2$ and $3$ qubit gates, corresponding to $C=5$ and $9$ control functions to be learnt respectively. 
For each problem the family of target $U^{tgt}_{\alpha}$ considered and the domain $\chi$ of the target parameters $\paramstgt$ are reported.
Results are provided in terms of the average (and standard deviations in bracket) infidelities $\bar{\mathcal{I}}$, and of the ratio $R$ between the average gate times resulting from a decomposition in terms of elementary gates and the times necessitated by the framework. Additional elements of training the families (vi) and (vii) are provided in the main text.}
\centering
\begin{tabular}{ll lcc} 
\hline\hline               
&$U^{tgt}_{\alpha}$   & \, \, $\chi$& $\bar{\cal I}[10^{-4}]$ & $R$ \\ [0.5ex]   
\hline
(i)	& $\ket{0}\bra{0}\otimes I+\ket{1}\bra{1}\otimes \exp(-i\alpha_1\sigma^{(2)}_z)$					& $[0, \pi]$			& $1[1]$	& $2.0$  \\
(ii)	& $\exp\left(-i\alpha_1 \sigma^{(1)}_z\sigma^{(2)}_z\right)$											&$[0, \frac{\pi}{2}]$		& $0[0]$	& $2.0$\\
(iii)	& $\ket{0}\bra{0}\otimes I+\ket{1}\bra{1}\otimes U^{1}_{\paramstgt}$				& $[0, \pi]^3$			& $3[4]$	& $2.1$\\
(iv)	& 
$\exp(-i\sum_{j\in \{x,y,z\}}\alpha_j\sigma_j^{(1)}\sigma_j^{(2)})$ & $[0, \frac{\pi}{2}]^3$	& $4[4]$	& $2.6$\\
(v)	& $\exp(-i \alpha_1 \sigma_z^{(1)}\sigma_z^{(2)}\sigma_z^{(3)})$					& $[0, \frac{\pi}{2}]$  		& $1[0]$	& $4.1$ \\
(vi)	& $\exp(-i\sum_{j\in \{x,y,z\}}\alpha_j\sigma_j^{(1)}\sigma_j^{(2)}\sigma_j^{(3)})$				& $[0, \frac{\pi}{2}]^3$ 	& $9[8]$	& $>10$\\
(vii)	& $\left(I-\ket{11}\bra{11}\right)\otimes I+\ket{11}\bra{11}\otimes U^{1}_{\paramstgt}$	& $[0, \pi]^3$ 	& $6[5]$	& $>10$\\
\hline 
\end{tabular}\label{tab:resmain}
\end{table}

Table~\ref{tab:resmain} summarizes the results for a few selected families of two- and three-qubit gates with the domain $\chi$ of the parameters $\alpha$ depicted in column $3$ and the obtained average infidelities (in multiples of $10^{-4}$) in column $4$ (further details of the NNs used are reported in Sec.~I~C of Supp. Matt.).
The two-qubit gates $(i)$ to $(iv)$ involve the optimization over $C=5$ control functions, and the three-qubit gates $(v)$ to $(vii)$ involve the optimization over $C=9$ control functions.

Consistently with the previous findings, low infidelities $\bar{\mathcal{I}}<5\times 10^{-4}$ are achieved for any of the families of two-qubit gates (i-iv) and for the one-dimensional family of three-qubit gates (v).

A straightforward application of the above framework to the problems $(vi)$ and $(vii)$ -- corresponding to $9$ time-dependent controls to be learnt and $3$-dimensional families to be realized -- however results in higher infidelities ($3[2.5]\times 10^{-3}$ for (vi) and $1.7[1.5] \times 10^{-3}$ for (vii)) than in the other cases.
Yet, the results of the optimizations contain clear indications towards steps to reach higher fidelities that are now further discussed.

First, the lowest fidelities are systematically obtained for values of $\alpha$ close to the boundary of its admissible domain $\chi$ (as can already be seen in Fig.~\ref{fig:1q_res_trainableT}(a)). Enlarging the range of values used for training by $20\%$ resolves this effect.
Second, the control functions identified as optimal have general properties that can be exploited to reduce the number of independent functions that need to be learnt.
In case (vi), the control solutions discovered by the framework satisfy the relation
$f_{\alpha}^{13}=f_{\alpha}^{1z}=f_{\alpha}^{3z}=0$ and in case (vii)
$f_{\alpha}^{1z}=f_{\alpha}^{2z}=0$,  $f_{\alpha}^{1y}=f_{\alpha}^{2y}$ with also $f_{\alpha}^{13}=f_{\alpha}^{23}$.
This indicates that only $6$ and $5$ independent control functions, out of the $9$ possible appearing in Eq.~\eqref{eq:H_nq}, are needed for the cases $(vi)$ and $(vii)$ respectively.
The infidelities listed in Table~\ref{tab:resmain}, for families $(vi)$ and $(vii)$, result from an optimization with enlarged domain $\chi$ and reduced number of control functions,
and their magnitude is comparable to those of the other cases.

While generally the non-uniqueness of solutions of optimal control problems makes it difficult to understand why a solution returned by a specific algorithm does achieve the goal that it is meant to achieve,
it seems that the requirement of smooth dependence on the parameters $\alpha$ helps the neural network to identify common features of all control pulses within the family and to avoid unnecessary terms in the Hamiltonian that would obscure its working principle.

Beyond this conceptual benefit and the small infidelities that are achieved, the gain in gate time is also of high practical relevance.
Since state-of-the-art implementation of unitaries on quantum devices rely on their decompositions in terms of elementary gates, the times $T^{dec}$ entailed by such decompositions provide well-defined baselines.
Given the freedom offered by the controlled Hamiltonian in Eq.~\eqref{eq:H_nq} these decomposition are performed in terms of the gate-set of rotations generated by the single qubit $\sigma_y$ and $\sigma_z$ and two-qubit $\sigma_x\sigma_x$ operators, for which qiskit's~\cite{Qiskit} compiling routine is employed with the highest level of optimization available (Sec.~II of Supp. Matt.).

Column $5$ of Table~\ref{tab:resmain} depicts the ratio $R$ between the averaged durations $\langle T^{dec}_\alpha\rangle_\alpha$ obtained with compiled gate circuits and the durations obtained with the present techniques.
In all cases there is an improvement of at least a factor of $2$, but, in cases $(vi)$ and $(vii)$, the improvement is substantially larger.
This suggests that compilation techniques (\ie discrete optimizations) struggle with these complex three-qubit gates, whereas the continuous optimization realized in terms of neural networks does not suffer from these limitations.

The ability to accurately control entire families of gates in reduced time, especially for complex gates, highlights the benefits of the present methodology.
Given that the automatic differentiation techniques~\cite{NEURIPS2018_69386f6b}, that ensures the efficient training of the framework,  can be applied to any system of ordinary differential equations,
\gqoc\ can find direct application to a broad range of quantum systems, such as superconducting qubits.
Since those are non-linear oscillators with a ladder of excited states, further studies would include suppression of leakage to such states.
Similarly, trapped ions or opto-mechanical systems with several interacting degrees of freedom pose control problems that can be addressed with the present techniques.
Additionally, automatic differentiation has also been extended to the treatment of stochastic differential equations (\eg~\cite{NEURIPS2019_59b1deff,rackauckas2020universal}) such that \gqoc\ can even generalize to problems of control with active feedback~\cite{schafer2021control,porotti2021deep}.

While optimal control is traditionally realized in terms of control pulses designed in numerical experiments, fundamental limitations in modelling and simulating the dynamics of composite quantum systems resulted in a shift towards designing control pulses in laboratory experiments~\cite{PhysRevLett.112.240504,Werninghaus2021,ZeroFid,baum2021experimental}.
Just like many techniques for individual control targets could be generalized to this setting, also \gqoc\ could be trained based exclusively on experimental data,
either in situations where gradients can be experimentally estimated~\cite{banchi2021measuring}, or by resorting to gradient-free optimization strategies~\cite{salimans2017evolution}. 

Essentially, the methodology that was presented here enables the control of a quantum system in different \emph{contexts}.
In the examples investigated, this context was in one to one correspondence with the target gate to be realized, that is, the overall details of the system under control were kept fixed and only the targets were varied.
More generally, the scheme based on NNs allows to tailor controls to be applied to any relevant context variable.
For instance, the inputs of the NN could also include intrinsic details of the controlled system (such as varied energy detunings~\cite{Porotti2019} or sizes~\cite{vanFrank2016}) or extrinsic (such as environmental heating rates~\cite{Horn_2018} or nearby operations inducing cross-talk~\cite{PhysRevLett.126.230502}).
Provided that the effects of these context variables can be simulated and that the corresponding optimal controls are expected to vary continuously with these variables, one would learn to accurately operate a quantum device in very broad situations.

We are indebted to stimulating discussions with Selwyn Simsek. This work is supported through a studentship in the Quantum Systems Engineering Skills and Training Hub at Imperial College
London funded by EPSRC (Grant No. EP/P510257/1),
and through  funding from the QuantERA ERANET Cofund
in Quantum Technologies implemented within the
European Union’s Horizon 2020 Programme under the
project Theory-Blind Quantum Control TheBlinQC and
from EPSRC under the grant EP/R044082/1.

\end{document}

\title{
Supplemental Material to Optimal control of families of quantum gates
}
\author{Fr\'ed\'eric Sauvage}
\author{Florian Mintert}

\affiliation{Physics Department, Blackett Laboratory, Imperial College London, Prince Consort Road, SW7 2BW, United Kingdom}
\date{\today}

\maketitle
The Supplemental Material is structured as follows.  In Sec.~\ref{sec:impl}, implementation details of the framework are provided.
In Sec.~\ref{sec:dec} we review the methodology permitting the systematic decomposition of arbitrary unitaries in terms of the elementary gates that are supported by the Hamiltonian in Eq.~(1), and the assessment of the resulting decomposition times. 
These times are used as a baseline for the ratios $R$ which are provided in Table~I.

\section{Implementation details}
\label{sec:impl}
Sec.~\ref{sec:ad} details the technical aspects of training the framework by means of gradient-descent and the numerical libraries that have been employed. 
Sec.~\ref{sec:withtime} reviews the extensions of the framework enabling the optimization of both the control functions and control times.
Finally, in Sec.~\ref{sec:hp} we describe how the hyperparameters of the framework are selected and provide details of the NNs which have been used for the results reported in Table~I.

\subsection{Training with gradient descent using the automatic differentiation toolbox}
\label{sec:ad}
Recall from the main text that, the training of the framework consists in optimizing the weights of the neural network (NN), denoted $\phi_{NN}$, to produce controls $\gcontrolfunc$ which minimize the deviation $\mathcal{I}$ between the induced propagators $U_{\alpha}$ and the targeted ones $U^{tgt}_{\alpha}$.
$\mathcal{I}$ is characterized by the average of the individual infidelities $\mathcal{I}_{\paramstgt}$ (Eq.~(2)), each corresponding to a fixed set of target parameters $\alpha$, taken over the domain of $\alpha$. 
Optimization of $\phi_{NN}$ to minimize $\mathcal{I}$ is performed by gradient-descent, and we now describe the steps of implementation required.

For fixed parameters $\paramstgt$ but variable time $t$, the NN (as sketched in Fig.~1) outputs the time-dependent control values $\gcontrolfunc$ used to construct the system Hamiltonian $H_\alpha(t)$ (\eg Eq.~(1)).
Dynamical evolution of the system under this Hamiltonian is performed by numerically solving the corresponding ordinary differential equation (ODE), which for closed quantum systems is the Schr\"{o}dinger equation.
Finally, the individual infidelity $\mathcal{I}_{\paramstgt}$ is evaluated based on the propagator $U_{\paramstgt} = U_{\paramstgt}(t = T)$, obtained at the end of the numerical evolution of the system, and its corresponding target $U_{\paramstgt}^{tgt}$.

Having detailed how $\mathcal{I}_{\alpha}$ is evaluated, it remains to detail how its gradients $\nabla_{\phi_{NN}}\mathcal{I}_{\paramstgt}$, with respects to the weights of the NN, are obtained.
Evaluations of these gradients necessitate to combine intermediate gradients resulting from each of the steps of computation discussed in the previous paragraph.
To make such gradient composition as effortless as possible, we resort to the automatic differentiation (AD) toolbox. 
In a nutshell, AD provides the machinery to systematically extend arbitrary numerical computations to simultaneously compute its derivatives, at the cost of only a modest computational overhead~\cite{BARTHOLOMEWBIGGS2000171,baydin2018automatic}.
This concept is central to the training of many machine learning algorithms, where it is most famously known as back-propagation.
Still, AD is not limited to the training of NNs but applies to virtually any computational routines.
In particular, recent advances in AD popularized in~\cite{NEURIPS2018_69386f6b} permit the efficient differentiation over the trajectory of a system which dynamics is obtained by numerically solving an ODE.
For the implementation of these techniques we resort to the package~\cite{torchdiffeq} built on top of the widely-adopted machine learning library pytorch~\cite{NEURIPS2019_9015}. 
Equipped with these libraries, it is only required to define how $\mathcal{I}_{\paramstgt}$ is evaluated (as was detailed in the previous paragraph), but, it is left to the underlying numerical routines to manage the intricacies of gradient computation.

Finally, the individual infidelities and corresponding gradients need to be averaged over the continuous domain of $\alpha$.
This average is approximated by means of Monte Carlo sampling over a finite number of $128$ values of $\alpha$ randomly drawn. 
Given the stochastic nature of the resulting gradients, we resort to the stochastic gradient optimization routine Adam~\cite{kingma2014adam}.

\subsection{Learning control functions and control durations}
\label{sec:withtime}
In addition to the derivatives with respect to the control values, the techniques in~\cite{NEURIPS2018_69386f6b} (see Appendix.C) can also evaluate derivatives $\nabla_T\mathcal{I}_{\alpha}$ of the infidelities with respect to the duration $T$ of the control tasks. 
Hence, this duration can be treated as a variable to be optimized over, rather than a fixed quantity.
Optimizing the cost $\mathcal{I_{\paramstgt}} + \mu \times T$, which now includes the penalty term $\mu \times T$ (with $\mu > 0$),
permits us to minimize the duration $T$ common to all the gates to be realized. This methodology is followed for the results presented in Table~I.

Going further, it is also possible to consider durations $T_\alpha$ which depend on $\alpha$, that is, target-dependent durations. 
Given that the exact nature of this dependency is unknown beforehand it needs to be learnt.
To this intent, a second NN, taking $\alpha$ as an input and returning $T_{\paramstgt}$ as an output, is trained concurrently with the original NN to minimize the cost $\mathcal{I_{\paramstgt}} + \mu \times T_{\paramstgt}$. 
Results of this approach were illustrated in Fig.~3.

In both these scenarios, the value of $\mu$ can be varied to put more or less emphasis on the objective of accurate control or on the objective of fast operations. 
The value of $\mu=10^{-2}$ was found to be a good compromise, and was used for the optimization results presented. 
Finally, given that the controls produced by the (original) NN should change depending on the control duration $T$ (or $T_{\alpha}$),
it is extended to also take the durations $T$ (or $T_{\paramstgt}$) as input. Note that this additional input is only required when the durations vary during the course of the optimization, that is, is not needed when $T$ is taken to be fixed (as was the case for the results presented in Fig.~2).

\subsection{Choice of the hyperparameters}
\label{sec:hp}
It is known that the quality of the solutions found by a NN can be affected by several design choices commonly referred to as hyperparameters~\cite{claesen2015hyperparameter}. 
In particular, the choice of the size of the NN, the way its weights are initialized, and the learning rates employed during its optimization are often considered to be of particular importance.
As the optimal hyperparameters values are not known beforehand, they need to be identified by means of trial-and-error. 
For instance, it is common to randomly sample~\cite{bergstra2012random} several hyperparameters configurations and to select a posteriori the ones resulting in optimal training. As each of these random configurations is evaluated independently, this approach has the benefit of being straightforward to parallelize.
To accelerate this search, it is based on partial training of the NNs with only a limited number $\sim 10$ of iterations, such as to quickly discover the most promising hyperparameters values. 
Once identified, these configurations are used to perform a complete training of the framework, consisting of a number of $400$ iterations reported in the main text.

The hyperparameters used for the results of Table.~I are reported in Table.~\ref{tab:hp}. 
The NNs employed are systematically taken to contain a number $n_L \in [4,10]$ of layers with, $n_N\in[150, 300]$ nodes per layer. 
The inner layers are chosen to be ReLU layers which are known to limit issues of gradient vanishing when training NNs~\cite{pmlr-v15-glorot11a}. 
The output layer, however, is taken to be a sigmoid layer which has bounded output values in the range $[0, 1]$, that are further shifted and rescaled to be in the desired range $[-1, 1]$ allowed for the control amplitudes. 
Learning rates used for the optimization of the weights of the NNs with Adam, are considered in the range $l\in [10^{-4} , 10^{-2}]$, a common choice when training NNs. 
Finally, we find the amplitudes of the initial weights of the NN to play an important role in the successful training of the framework. In particular, the default distribution of the initial weights yields small initial control amplitudes, which are prone to quickly converge to $0$. To avoid such effect, the initial weights are rescaled by a factor $\beta \in [1.8, 2.2]$ such that the control values initially produced by the NNs are different enough from $0$.

\setlength{\tabcolsep}{5pt}
\begin{table}[t]
\caption{Details of the hyperparameters used for the $7$  families of unitaries studied in Table.~I.
These include the number $n_L$ of NN layers and the number $n_U$ of units per layer, the learning rate $l$ of the Adam optimizer and the scales $\beta$ of the initial weights of the NN.
} 
\centering
\begin{tabular}{ll ccc} 
\hline\hline               
& $U^{tgt}_{\alpha}$  & $n_L \times n_U$ & $l (10^{-3})$ & $\beta$\\    
\hline
(i)	& $\ket{0}\bra{0}\otimes I+\ket{1}\bra{1}\otimes \exp(-i\alpha_1\sigma^{(2)}_z)$ & $6 \times 150$ & $1$ & $1.8$ \\
(ii)	& $\exp\left(-i\alpha_1 \sigma^{(1)}_z\sigma^{(2)}_z\right)$ & $6 \times 150$ & $1$  & $2.1$ \\
(iii)	& $\ket{0}\bra{0}\otimes I+\ket{1}\bra{1}\otimes U_{1}(\vec\paramstgt)$ & $8 \times 300$ & $3$  & $1.9$\\
(iv)	& 
$\exp(-i\sum_{j}\alpha_j\sigma_j^{(1)}\sigma_j^{(2)})$ & $8 \times 300$ & $1$  & $1.9$\\
(v)	& $\exp(-i \alpha_1 \sigma_z^{(1)}\sigma_z^{(2)}\sigma_z^{(3)})$ & $10 \times 200$ & $4$  & $2.2$\\
(vi)	& $\exp(-i\sum_{j}\alpha_j\sigma_j^{(1)}\sigma_j^{(2)}\sigma_j^{(3)})$ & $10 \times 300$ & $2$  & $2.0$\\
(vii)	& $\left(I-\ket{11}\bra{11}\right)\otimes I+\ket{11}\bra{11}\otimes U_{1}(\vec\paramstgt)$ & $10 \times 300$ & $4$   & $2.1$\\
\hline 
\end{tabular}\label{tab:hp}
\end{table}

\section{Evaluation of the times entailed by gate decomposition}
\label{sec:dec}
In Table~I are reported the ratios $R$ between the times $T^{dec}$ entailed by gate decomposition of the unitaries considered and the control times $T$ required by the framework. We now detail the methodology used to evaluate $T^{dec}$.

Gate decompositions are performed with respect to the set of elementary gates which are supported by the Hamiltonian in Eq.~(1), that is, the gate set
\begin{equation}
    \mathcal{G}=\{R_\sigma(\theta)=\exp(-i \frac{\theta}{2} \sigma)\}
\end{equation} 
of rotations generated by the operators $\sigma \in \{\sigma^{(i)}_y,\sigma^{(i)}_z, \sigma^{(i,i'\neq i)}_{xx}\}$ acting on qubits $i,i'=1,\hdots,n$.
To systematize such decompositions, we rely on established automated routines and resort to the qiskit compilation algorithm. 
Among the gate sets supported by qiskit, we select the gateset composed of two qubits rotations $R_{\sigma_{xx}}$ (between arbitrary pairs of qubits) and single qubit rotations 
\begin{equation}
U_3(\theta, \phi, \lambda) = \begin{pmatrix}
\cos (\frac{\theta}{2}) &  - e^{i \lambda} \sin (\frac{\theta}{2})\\
e^{i\phi} \sin (\frac{\theta}{2}) & e^{i (\lambda+\phi)} \cos (\frac{\theta}{2})
\end{pmatrix},
\end{equation}
which can be further broken down as elements of $\mathcal{G}$ as $R_{\sigma_z}(-\lambda)R_{\sigma_y}(\theta)R_{\sigma_z}(-\phi)$, up to a global phase.

Given the resulting gate decomposition of a unitary $U^{tgt}_{\alpha}$, 
it remains to assess the time $T^{dec}_{\alpha}$ that would be needed for its implementation.
For the range of values $[-1,1]$ permitted for the controls of the Hamiltonian in Eq.~(1), it follows that the minimal time required for the implementation of any gate $R_{\sigma}(\theta)$ is $|\theta / 2|$. 
Hence, a sequential application of $l$ gates $\prod_l R_{\sigma_l}(\theta_l)$ necessitates an implementation time of $T^{dec}=\sum_l |\theta_l / 2|$.
For instance, a gate $U_3(\theta, \psi, \lambda)$ requires a time $T^{dec}=(|\theta| + |\psi| +|\lambda|)/2$.
It should be noted that distinct sets of angles $(\theta, \psi, \lambda)$ can yield the same unitary $U_3(\theta, \psi, \lambda)$ but could entail different implementation times; the fastest implementation is systematically considered. 
Finally, the times $T^{dec}_{\alpha}$ effectively depend on the individual targets $U^{tgt}_{\paramstgt}$ to be realized. As such, for each of the $7$ families of gates presented in Table.~I, these times are averaged over an ensemble of $100$ gates randomly drawn.